\documentclass[aps,prl,reprint,floatfix,preprintnumbers,showpacs,amsfonts,nofootinbib,superscriptaddress]{revtex4-1}
\usepackage{graphicx}
\usepackage{amssymb,amsmath}
\usepackage{color}

\newcommand*{\bq}{\mathbf{q}}
\newcommand*{\chiimp}{\chi_{\text{imp}}}

\newcommand*{\pdag}{\phantom{\dag}}

\begin{document}

\title{Quantum phase transitions into Kondo states in bilayer graphene}
\date{\today}

\author{Diego Mastrogiuseppe}
\affiliation{Department of Physics and Astronomy, and Nanoscale
and Quantum Phenomena Institute, \\ Ohio University, Athens, Ohio
45701--2979}
\affiliation{Dahlem Center for Complex Quantum Systems and Fachbereich Physik,
Freie Universit\"at Berlin, 14195 Berlin, Germany}
\author{Arturo Wong}
\affiliation{Department of Physics, University of Florida, P.O.\ Box 118440,
Gainesville, Florida, 32611--8440}
\author{Kevin Ingersent}
\affiliation{Department of Physics, University of Florida, P.O.\ Box 118440,
Gainesville, Florida, 32611--8440}
\author{Sergio E.\ Ulloa}
\affiliation{Department of Physics and Astronomy, and Nanoscale
and Quantum Phenomena Institute, \\ Ohio University, Athens, Ohio
45701--2979}
\affiliation{Dahlem Center for Complex Quantum Systems and Fachbereich Physik,
Freie Universit\"at Berlin, 14195 Berlin, Germany}
\author{Nancy Sandler}
\affiliation{Department of Physics and Astronomy, and Nanoscale
and Quantum Phenomena Institute, \\ Ohio University, Athens, Ohio
45701--2979}
\affiliation{Dahlem Center for Complex Quantum Systems and Fachbereich Physik,
Freie Universit\"at Berlin, 14195 Berlin, Germany}

\begin{abstract}
We study a magnetic impurity intercalated in bilayer graphene. A representative
configuration generates a hybridization function with strong dependence on the
conduction-electron energy, including a full gap with one hard and one soft
edge. Shifts of the chemical potential via gating or doping drive the system
between non-Kondo (free-moment) and Kondo-screened phases, with strong
variation of the Kondo scale. Quantum phase transitions near the soft edge
are of Kosterlitz-Thouless type, while others are first order. Near the hard
edge, a bound-state singlet appears inside the gap; although of single-particle
character, its signatures in scanning tunneling spectroscopy are very similar
to those arising from a many-body Kondo resonance.
\end{abstract}


\maketitle

One of the remarkable manifestations of cooperative phenomena in condensed
matter is the many-body screening of a magnetic impurity in a nonmagnetic
metal. This Kondo effect, well understood in ordinary metals \cite{Hewson},
acquires added complexity in cases where the host density of states (DOS)
varies strongly with energy $E$ near the chemical potential $\mu$.
The pseudogap Kondo problem \cite{WF90, BPH97, GI98, GL03, FV04} with a DOS
$\rho(E)\propto|E-\mu|^r$ (realized, for example, for $r=1$ in
high-temperature superconductors \cite{VB01}) exhibits a rich phase
diagram that depends on the band exponent $r$, the impurity-host exchange
coupling $J$, and the presence or absence of particle-hole (p-h) symmetry.

Similar DOS features can also appear in low-dimensional systems such as
graphene \cite{CGP09}. The technologically and conceptually important issue of
creating localized magnetic moments in monolayer graphene, and the
appearance of the Kondo effect, have been the focus of many recent theoretical
\cite{HG07,SB08,CUB09,WBK10,ZDB10,VFB10,JK11,CA11,URC11,KMO12,CIG12,KK13} and
experimental \cite{BDS10, manoharan, CLC11, HZW12} studies, with controversial
results (see \cite{FV13} for an overview).
A more complex DOS appears in bilayer graphene (BLG), a material that can be
gapped by gating, and thus has attracted much attention for possible device
applications \cite{CNM07}. The variety of microscopic environments for magnetic
impurities combined with easy tunability, make BLG highly promising for the
study of quantum phase transitions (QPTs) into various Kondo states
\cite{DZB09, KHP11}.

This paper explores such QPTs for a representative configuration of an
intercalated spin $\sigma=1/2$ magnetic impurity in BLG. This setup is described by an
Anderson impurity model with an energy-dependent hybridization featuring a gap
that has one hard and one soft edge.
Under variation of the chemical potential $\mu$, the system passes from a
free-moment (FM) phase to a Kondo phase featuring a strong $\mu$ dependence of
the Kondo temperature scale. The QPTs found near the soft hybridization edge
are of Kosterlitz-Thouless type, while all other QPTs are first order. We
present thermodynamic and spectral properties near these QPTs, and discuss
some of their consequences for spectroscopy measurements. 
For $\mu$ near the hard hybridization edge, the FM phase exhibits a
singlet bound state inside the gap. This bound state is of single-particle
character, but may have signatures in scanning tunneling spectroscopy very
similar to those arising from a many-body Kondo resonance.
Many of these features derive from the nature of the underlying DOS and are
shared by other impurity configurations. More importantly, as the chemical
potential can be effectively shifted by symmetric gating or homogeneous doping
of the sample, the predicted properties should be accessible in available
experimental setups, enabling detailed exploration of the physics of QPTs.

\begin{figure}[t]
\includegraphics[width=0.8\columnwidth]{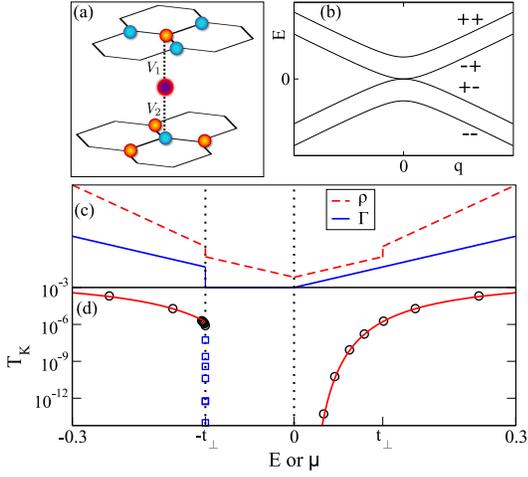}
\caption{(Color online)
(a) Lattice structure of Bernal-stacked BLG, showing impurity
configuration considered here.
(b) Dispersion $E_{\alpha\alpha'}(q)$ of the four bands along a line
through one of the Brillouin-zone corners $K_{\pm}$ (where $q=0$).
(c) BLG density of states $\rho(E)$ (dashed line) and 
hybdridization function $\Gamma(E)$ for the impurity configuration in (a) with
$V_1=V_2$ (solid line).
(d) Kondo temperature $T_K$ for $U=-2\epsilon_d=0.1$ and $V_1=V_2=0.21$ vs
chemical potential $\mu$ outside (circles) and inside (squares) the
hybridization gap. The (red) line shows the result of substituting $\Gamma(\mu)$
into Haldane's formula for $T_K$.}
\label{fig1}
\end{figure}

Bernal-stacked BLG [see Fig.\ \ref{fig1}(a)] is modeled with a
real-space tight-binding Hamiltonian \cite{MF06}
with nearest-neighbor intra-layer hopping amplitude $t$ 
and inter-layer hopping $t_{\perp}\simeq 0.1t$ between A-sublattice sites in
layer 1 and B-sublattice sites in layer 2.
After Fourier transformation, this Hamiltonian can be expanded around
inequivalent Brillouin zone corners $K_{\pm}$ that form the centers of two
valleys $\tau=\pm$ in the gapless spectrum of neutral BLG. Diagonalization of
the $4\times4$ Hamiltonian describing each valley \cite{MF06,suppl} yields
hyperbolic bands labeled $\alpha,\,\alpha'=\pm$ [see Fig.\ \ref{fig1}(b)]
having dispersion $E_{\alpha,\alpha'}(q) = \alpha\,t_{\perp}/2
+ \alpha'\sqrt{(v_F q)^2+(t_{\perp}/2)^2}$, where $q$ is the magnitude of the
wave vector $\bq$ measured relative to the zone corner and $v_F=3at/2$ is the
Fermi velocity of graphene. The combined DOS of both valleys (per spin
orientation, per unit cell of area $\Omega_0$),
\begin{equation}
\rho(E) = \frac{\Omega_0}{\pi v_F^2} \biggl[ |E| + \frac{t_{\perp}}{2} +
  \Theta(|E|-t_{\perp}) \biggl(|E| - \frac{t_{\perp}}{2} \biggr) \biggr],
\end{equation}
has jumps at $E=\pm t_{\perp}$ [dashed line in Fig.\ \ref{fig1}(c)].
When its layers are gated symmetrically to yield a common chemical potential
$\mu$ or homogeneously doped via intercalates, 
BLG is described by an eight-band low-energy effective Hamiltonian (per spin)
$H_{\text{BLG}} =
\sum_{\alpha,\alpha',\tau,\sigma,\bq}$ $[E_{\alpha,\alpha'}(q)-\mu]$
$c_{\alpha,\alpha',\tau,\sigma}^{\dag}(\bq)$
$c_{\alpha,\alpha',\tau,\sigma}^{\pdag}(\bq)$. We take $t=3.0$\,eV,
$t_{\perp}=0.3$\,eV, and $a=1.42$\,\AA, in which case $v_F\simeq 10^6$\,m/s
and $H_{\text{BLG}}$ has an effective half-bandwidth $D\simeq 2.5$\,eV,
which we take as the energy unit in our calculations.

An intercalated impurity can occupy one of several inequivalent positions
\cite{KKS11}. We focus on the high-symmetry configuration shown in Fig.\
\ref{fig1}(b), described by an Anderson Hamiltonian
$\bar{H}_A=H_{\textit{BLG}}+\epsilon_d n_d + U n_{d\uparrow}\,n_{d\downarrow}
+ N_c^{-1/2} \sum_{\sigma} \{[ V_1 a_{1,\sigma}^{\dag}(\mathbf{0})
+ V_2 b_{2,\sigma}^{\dag}(\mathbf{0}) ] d_{\sigma}^{\pdag}$ $+\text{H.c.}\}$,
where $n_d=n_{d\uparrow} + n_{d\downarrow}$ with
$n_{d\sigma} = d_{\sigma}^{\dag} d_{\sigma}^{\pdag}$ being the impurity number
operator for spin $\sigma=\pm1/2$, $\epsilon_d$ is the impurity level energy relative
to the chemical potential, $U$ is the local Coulomb repulsion, and $N_c$ is
the number of BLG unit cells. The last term in $\bar{H}_A$ describes tunneling
of an electron between the impurity and the nearest sublattice-A atom in
layer 1 or the nearest B-sublattice atom in layer 2. Transformation to the
eigenbasis of $H_{\mathrm{BLG}}$ and thence to an energy representation yields
\cite{suppl}
\begin{align}
\label{H_A}
H_A
&= \sum_{\sigma}\int_{-D}^D dE\;(E-\mu)\,c_{E\sigma}^{\dag} c_{E\sigma}^{\pdag}
  + \epsilon_d n_d + U n_{d\uparrow} \, n_{d\downarrow} \notag \\[-1ex]
&+ \sum_{\sigma} \int_{-D}^D dE\,\sqrt{2\Gamma(E)/\pi} \:
  \Bigl( c_{E\sigma}^{\dag} \, d_{\sigma}^{\pdag} + \text{H.c.} \Bigr) .
\end{align}
Here $c_{E\sigma}^{\dag}$, satisfying
$\{ c_{E\sigma}^{\dag}, \, c_{E'\sigma'}^{\pdag} \}
= \delta(E\!-\!E') \, \delta_{\sigma,\sigma'}$, creates an electron in the
single linear combination of band states of energy $E$ that hybridizes with the
impurity, and we have dropped contributions from all the other (decoupled)
linear combinations of band states. $H_A$ represents a conventional Anderson
model apart from the unusual energy dependence of the hybridization function
\begin{equation}
\label{Gamma}
\Gamma(E) = \frac{\Omega_0 |E|}{8 v_F^2} \sum_{\alpha=\pm}
    (V_2 - \alpha V_1)^2 \, [\Theta(-\alpha E)+\Theta(\alpha E-t_\perp)] .
\end{equation}

The relative contributions of the eight BLG bands to $\Gamma(E)$ depend on the
ratio of tunneling amplitudes $V_1/V_2$. Figure \ref{fig1}(c) sketches the
variation of $\Gamma(E)$ for $V_1=V_2$. 
The coupling of the impurity to BLG breaks particle-hole symmetry in general, 
so that the symmetry of $\rho(E)$ under $E\to-E$ is broken
in $\Gamma(E)$, which has a gap with a jump onset at its lower edge
($E=-t_{\perp}$) and a linear-in-energy onset at the upper edge ($E=0$).  The
gap in the hybridization function can be traced back to the symmetries of the 
band-states under inversion through the BLG plane \cite{suppl}.
By analogy with the behavior a magnetic impurity in a host with a hard gap
\cite{CC98, GL08a, GL08b, MR10} or a power-law pseudogap \cite{WF90, BPH97,
GI98, GL03, FV04}, one expects this form of $\Gamma(E)$ to produce a
free-moment (\textbf{FM}) phase spanning the parameter range
$-t_{\perp}\lesssim \mu\lesssim 0$, $-U\lesssim\epsilon_d\lesssim0$,
surrounded by a strong-coupling (\textbf{SC}) phase in which the impurity
moment is fully quenched at temperature $T=0$.

For $V_1 = -V_2$, the hybridization in Fig.\ \ref{fig1}(c) must be reflected
about $E=0$, but otherwise the physics is the same as for $V_1=V_2$. In cases
$|V_1| \ne |V_2|$ that may arise if the impurity sits closer to one graphene
layer than the other, Eq.\ \eqref{Gamma} shows that $\Gamma(E)=0$ only at
$E=0$.
Nonetheless, if $V_1$ and $V_2$ have similar magnitudes, there will be a range of
chemical potentials just above or below $\mu=0$ within which $\Gamma(\mu)$ is
so small that any Kondo screening takes place below experimentally accessible
temperatures, and the measured properties will be indistinguishable from those
for $|V_1|=|V_2|$.

To substantiate the picture outlined above, we have studied the case $V_1=V_2$
using the numerical renormalization-group (NRG) \cite{BCP08}, a nonperturbative
method that allows $H_A$ to be diagonalized iteratively to obtain the low-lying
many-body states, which can be used to calculate the impurity occupancy
$\langle n_d\rangle$, its contribution to the static magnetic
susceptibility $\chiimp(T)$, and the impurity spectral function $A_d(\omega,T)$.
We adopt units where $\hbar=k_B=g\mu_B=D=1$. 
All results shown are for $U=0.1$
and $V_1=V_2=0.21$, calculated with an NRG discretization parameter $\Lambda=2.5$
and retaining 2\,000 many-body states (each one representing a degenerate spin
multiplet) after each iteration.

\begin{figure}[tb]
\includegraphics[width=0.8\columnwidth]{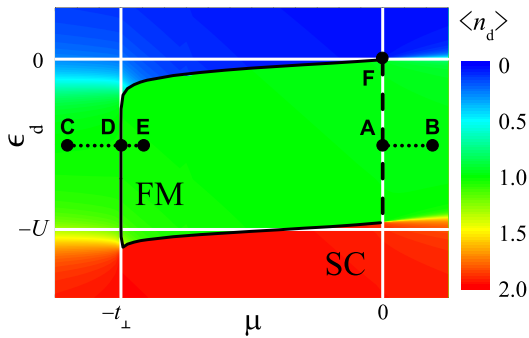}
\caption{(Color online) Ground-state impurity occupancy $\langle n_d\rangle$ on
the $\mu$-$\epsilon_d$ plane for $U=0.1$ and $V_1=V_2=0.21$. Thick black lines
demarcate free moment (FM) and strong coupling (SC) phases. The occupancy
varies continuously across a line of Kosterlitz-Thouless QPTs at $\mu=0$
(dashed). Elsewhere around the phase boundary, $\langle n_d\rangle$ jumps
across a first-order line (solid). White lines at $\mu=0,\,-t_{\perp}$ delimit
the hybridization gap; that at $\epsilon_d=0$ ($-U)$ shows where
$\langle n_d\rangle$ would jump from 0 to 1 (1 to 2)
for $V_1=V_2=0$. \textsf{A} through \textsf{F} schematically
represent endpoints of paths discussed in the text.}
\label{fig2}
\end{figure}

We first consider the global phase diagram. Figure \ref{fig2} maps the $T=0$
impurity occupancy on the $\mu$-$\epsilon_d$ plane. There are two phases,
within each of which $\langle n_d\rangle$ varies smoothly under change of $\mu$
and/or $\epsilon_d$.
An FM phase spans a contiguous region (bounded by thick lines in Fig.\
\ref{fig2}) that largely coincides with the one (inside white lines) in which
$\Gamma(\mu)=0$ \textit{and} the impurity level would be singly occupied in the
atomic limit $V_1=V_2=0$. Throughout most of the FM phase,
$0.9 < \langle n_d\rangle < 1.1$, although strong departures from this range
occur in the lower-left and upper-right corners. The rest of the plane
is taken up by an SC phase. Around three sides (solid line) $\langle n_d\rangle$
jumps on crossing the phase boundary, with $|\Delta\langle n_d\rangle|$
exceeding 0.9 along most of the top and bottom sides but generally being smaller
than $0.01$ along a near-vertical section at $\mu\simeq -t_{\perp}$. By contrast,
the occupancy varies smoothly across the boundary at $\mu=0$ (dashed line), where
the existence of a QPT is seen in properties other than $\langle n_d\rangle$.
The SC (green) region for $\mu < -t_\perp$ and $\mu > 0$ would result in 
strong Kondo signatures in STM experiments. 

\begin{figure}[t]
\includegraphics[width=0.95\columnwidth]{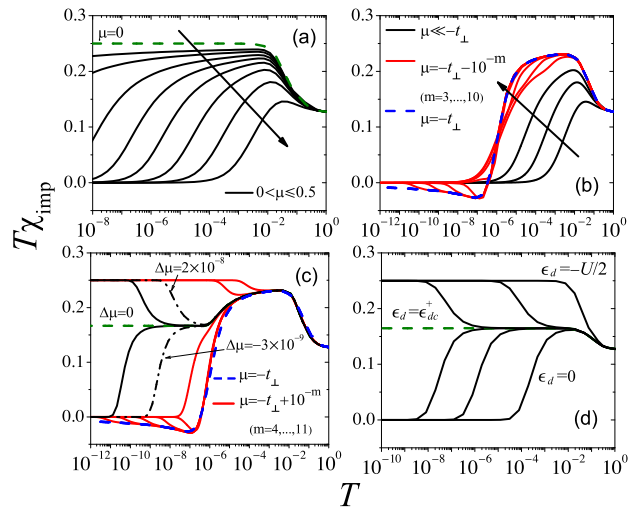}
\caption{(Color online) (a)--(c) Impurity susceptibility $T\chiimp(T)$
for $\epsilon_d=-U/2$ and different values of $\mu$ (increasing in the
direction of the arrows) in the ranges (a) $0 \le\mu\le 0.5$,
(b) $\mu\le -t_{\perp}$, and (c) $-t_{\perp}\le\mu<-t_{\perp}+10^{-4}$.
In (c), $\Delta\mu=\mu-\mu_c^-$ with
$\mu_c^- \simeq -t_{\perp}+1.16\times 10^{-6}$. 
(d) $T\chiimp(T)$ for $\mu=0$ and different values of $\epsilon_d$ between
$-U/2$ and $0$, with $\epsilon_{dc}^+\simeq -3.22\times 10^{-4}$.
Notice $t_\perp=0.12$.}
\label{fig3}
\end{figure}

We now examine the magnetic susceptibility along various paths crossing the
phase boundary, beginning near the right edge. Figure \ref{fig3}(a) shows
$T\chiimp$ vs $T$ for $\epsilon_d=-U/2$ and a set of $\mu$ values rising from
0 (shown schematically as path \textsf{A}--\textsf{B} in Fig.\ \ref{fig2}).
For any $\mu > 0$, $T\chiimp$ vanishes as $T\to 0$, signaling the complete
screening of the impurity magnetic moment at an SC fixed point. The
susceptibility follows a scaling form $T\chiimp(T)=f(T/T_K)$ for $T\lesssim
T_K$ defined via $T_K\chiimp(T_K)=0.0701$, and satisfies the standard relation
$\chiimp(0)=0.103/T_K$ \cite{W75}. The Kondo temperature [circles in Fig.\
\ref{fig1}(d)] is well-captured by the Haldane formula \cite{H78} $T_K =
\sqrt{U\Gamma/2} \, \exp[\pi\epsilon_d(U+\epsilon_d)/2U\Gamma]$
[line in Fig.\ \ref{fig1}(d)] for a constant hybridization function given by
$\Gamma\equiv\Gamma(\mu)$.
The impurity spectral function (not shown) exhibits a Kondo resonance width
proportional to $T_K$, and Hubbard bands near $\omega = \pm U/2$. These
properties, all characteristic of a conventional Kondo effect \cite{W75},
indicate that in this range of chemical potentials the variation of
$\Gamma(E)$ away from $E=\mu$ has negligible effect on the Kondo physics.

As $\mu$ approaches zero from above, $T_K$ decreases rapidly, exhibiting an
exponential sensitivity to the vanishing of $\Gamma(\mu)$, as expected from
Haldane's formula. However, for $\mu=0$ (point \textsf{A} in Fig.\ \ref{fig2}),
$T\chiimp$ does not drop toward zero as $T\to 0$,
but rather rises to the value 1/4 produced by a free spin-$\frac{1}{2}$ moment.
For $\mu<0$, $T\chiimp$ rises above the $\mu=0$ curve for intermediate $T$ 
before approaching the asymptote of 1/4 (not shown). 
This flow to the FM fixed point
has no characteristic temperature scale analogous to $T_K$. The impurity
spectral function shows no features on scales $|\omega|\ll U/2$. These are all
properties of a Kosterlitz-Thouless (KT) QPT (similar to that found in
the conventional Anderson model in the limit of vanishing hybridization), an
interpretation consistent with the smooth evolution of $\langle n_d\rangle$
across the phase boundary. Calculations for other values of $\epsilon_d$ provide
evidence for a line of KT fixed points at $\mu_c^+(\epsilon_d)=0$.

Moving to the left edge of the FM phase, we next consider $\epsilon_d=-U/2$
and various $\mu$ values spanning $\mu=-t_{\perp}$ (shown schematically as
path \textsf{C}--\textsf{E} in Fig.\ \ref{fig2}). Figure \ref{fig3}(b) shows
that the $T\to 0$ behavior of $T\chiimp(T)$ remains conventional for
$\mu\lesssim -t_{\perp}-10^{-5}$. $T_K$ as defined by $T_K\chiimp(T_K)=0.0701$
is in good agreement with Haldane's formula [Fig.\ \ref{fig1}(d)]. By contrast,
for $|\mu + t_{\perp}|\lesssim 10^{-7}$ [Figs.\ \ref{fig3}(b) and \ref{fig3}(c)],
over which range $T_K$ is almost constant, $\chiimp$ changes sign at
$T\simeq T_K/2$ and then approaches zero from below. Such a sign change, seen
in other systems with strong variation of the hybridization near the chemical
potential \cite{HK99,DSI06,ZI11} but generally without associated QPTs , arises from the discontinuity in $\Gamma(E)$
at $E=-t_{\perp}$ \cite{suppl}. 

Figure \ref{fig3}(c) shows that upon a small increase in $\mu$ further above
$-t_{\perp}$, $T\chiimp$ reverts to approaching zero from above, but exhibits a
scaling $T\chiimp(T\ll T_K)=f(T/T_K)$ with a different $f$ than in
the conventional Kondo regimes $\mu>0$ and $\mu\lesssim -t_{\perp}-10^{-5}$.
Moreover, $T\chiimp$ remains on a plateau at $1/6$ down to ever-lower
temperatures as $\mu$ increases toward
$\mu_c^-(\epsilon_d=-U/2)\simeq -t_{\perp}+1.16\times 10^{-6}$,
where $T\chiimp=1/6$ persists to $T=0$. For $\mu>\mu_c^-$, $T\chiimp$ instead
rises from the plateau to reach its free-spin-$\frac{1}{2}$ value $T\chiimp=1/4$
as $T\to 0$. In this part of the FM phase, one may also define a crossover
temperature $T_X$ via the criterion $T_X\chiimp(T_X)=1/5$ (say). $T_K$ in the
SC phase and $T_X$ in the FM phase both vanish linearly with $\mu-\mu_c^-$.
These properties and the jump in $\langle n_d\rangle$ noted above point
to a first-order QPT arising from the crossing of FM doublet and SC singlet
ground states. Essentially the same behaviors are seen in the Anderson model
with a power-law pseudogap described by a superlinear energy exponent
\cite{GI98,FV04}.
Similar behavior to that for $\epsilon_d=-U/2$ occurs elsewhere along the left
edge of the FM phase with the value of $\mu_c^- + t_{\perp}$ varying with
$\epsilon_d$ but remaining small and positive. 

Moving round to the top and bottom portions of the phase boundary, one no
longer finds sign changes in $\chiimp$, but indications of a level-crossing
QPT extend all the way to the corners at $\mu=0$. Figure \ref{fig3}(d)
illustrates $T\chiimp(T)$ for $\mu=0$ and different level energies in the
range $0\ge\epsilon_d\ge-U/2=-0.05$ (path \textsf{A}--\textsf{F} in Fig.\
\ref{fig2}), showing that $T\chiimp=1/6$ persists to $T=0$ at
$\epsilon_{dc}^+(\mu=0)\simeq -3.22\times 10^{-4}$, representing the upper
end of the first-order line. A similar QPT (not shown) anchors the lower end at
$\epsilon_{dc}^-(\mu=0)\simeq -0.0959$.

\begin{figure}[t]
\includegraphics[width=0.95\columnwidth]{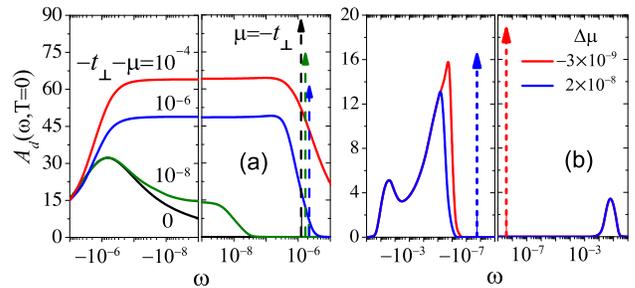}
\caption{(Color online)
Impurity spectral function $A_d(\omega,T=0)$ for $\epsilon_d=-U/2$ and
(a) four values of $\mu$ just below $-t_{\perp}$, and (b) two values of
$\mu$ straddling
$\mu_c^-$. Dashed lines denote poles inside the
hybridization gap, corresponding to singlet bound states.}
\label{fig4}
\end{figure}

Finally, we turn to the behavior of the $T=0$ impurity spectral function near
the left edge of the FM phase, as illustrated for $\epsilon_d=-U/2$ in Fig.\
\ref{fig4} (again, path \textsf{C}--\textsf{E} in Fig.\ \ref{fig2}).
Well into the Kondo regime, for $\mu+t_{\perp}\ll -T_K$ [see, e.g., the curve
for $\mu = -t_{\perp}-10^{-4}$ in Fig.\ \ref{fig4}(a)], the only low-energy
spectral feature is a Kondo resonance of width $T_K$ centered on
$\omega=0$. As $\mu$ gets closer to $-t_{\perp}$, $T_K$ remains
almost constant but the Kondo peak loses its spectral weight at frequencies
$\omega>-t_{\perp}-\mu$ lying inside the hybridization gap, leaving a weaker
virtual bound state resonance centered at $\omega\simeq -T_K$.
Simultaneously, a pole appears in $A_d(\omega,0)$ at a frequency
$\omega_b\simeq T_K$ inside the gap. This pole is associated with a bound state
produced by potential scattering from the impurity, a feature already present in
the noninteracting limit $U=0$ \cite{BalatskyRMP2006} (although rescaled for
$U>0$).

Once $\mu$ exceeds $-t_{\perp}$, the hybridization gap pushes the low-frequency
continuum portion of $A_d(\omega,T=0)$ into the range $\omega<-t_{\perp}-\mu$,
as illustrated in Fig.\ \ref{fig4}(b) for
$\Delta\mu = \mu-\mu_c^- = -3\times 10^{-9}$ and $2\times 10^{-8}$
[values of $\mu$ corresponding to dashed lines in Fig.\ \ref{fig3}(c)]. 
Just as for $\mu \lesssim -t_{\perp}$, a
bound state appears at a frequency $\omega_b\simeq T_K$. As $\mu$ increases
through $\mu_c^-$, the bound state passes smoothly through $\omega=0$ to
take up a \textit{negative} frequency $\omega_b\simeq -T_X$ in the FM phase
[see Fig.\ \ref{fig4}(b)], and switches character from a spin-singlet to a
doublet. Further increase of $\mu$ eventually causes the bound state to merge
into the continuum once $T_X\gtrsim t_{\perp}+\mu$.

The preceding results suggest that for $\mu$ sufficiently close to the
hybridization edge at $-t_{\perp}$, scanning tunneling microscopy should detect
a very sharp resonance close to the Fermi energy. Although the bound state
giving rise to this resonance is of single-particle character, it may be
difficult to distinguish from the many-body Kondo resonance that occurs
deeper into the SC phase.

This rich and complex behavior could be probed in experimentally accessible
systems such as BLG intercalated with nonmagnetic atoms (like Li \cite{KCU12},
which functions as electron dopant) or molecules, together with a low
concentration of magnetic impurities. A change in $\mu$ could be
achieved through gating as well, in a way that does not substantially modify
the original band structure \cite{KHM09}.

In summary, we have presented nonperturbative solutions of an Anderson model
describing a magnetic impurity intercalated in bilayer graphene (BLG).
In the high-symmetry impurity configuration considered, tunneling interference
effects combine with the BLG density of states to impart a strong energy
dependence to the impurity-host hybridization function. As a consequence, the
system exhibits both a Kondo-screened phase and a non-Kondo phase in which
an unquenched impurity moment survives to absolute zero. The phase boundary is
marked by lines of quantum phase transitions, some of which are first order
while the others are of Kosterlitz-Thouless type. For chemical potentials that
lie close to a jump onset in the hybridization function, single-particle
bound states may give rise to signatures in scanning tunneling microscopy very
similar to those produced by the many-body Kondo resonance.

This work was supported under NSF Materials World Network Grants
DMR-1107814 (Florida), and DMR-1108285 (Ohio), as well as by NSF-PIRE grant
0730257.
D.M., N.S., and S.E.U.\ acknowledge the hospitality of the Dahlem Center and
support from the A.\ von Humboldt Foundation.


\end{document}